\documentclass{ws-procs975x65}
\def\cN{{\cal N}}

\newcommand{\be}{\begin{equation}}
\newcommand{\ee}{\end{equation}}
\newcommand{\bea}{\begin{eqnarray}}
\newcommand{\eea}{\end{eqnarray}}
\newcommand{\ba}{\begin{array}}
\newcommand{\ea}{\end{array}}
\def\nn{\nonumber}

\newcommand{\Z}{{\mathbb Z}}

\begin{document}

\title{NON-GEOMETRICAL COMPACTIFICATIONS \\ WITH FEW MODULI}

\author{GIANFRANCO PRADISI}

\address{Dipartimento di Fisica and Sezione I.N.F.N.
\\
Universit\`a di Roma ``Tor Vergata''\\
Via della Ricerca Scientifica, 1\\00133 Roma, ITALY\\
E-mail: gianfranco.pradisi@roma2.infn.it}

\begin{abstract}
The stabilization of moduli is one of the main problems in string theory. In this talk I will
discuss some stringy mechanisms based on non-geometrical compactifications to obtain
four dimensional models with a reduced number of moduli.
\end{abstract}


\bodymatter

\section{Introduction}\label{aba:sec1}
Moduli stabilization is a crucial issue in order to compare string theory with phenomenology\cite{Antoniadis:2007uz}.  In the last few years, several mechanisms have been studied to realize it, related mainly to flux compactifications \cite{flux} and non-perturbative effects.  However, an exact quantization of string theory in the presence of fluxes is in general not accessible, excluding some very interesting cases of Type I models with open-string (oblique) magnetic fluxes\cite{as,ambt}.  The only possibility is to resort to the low-energy effective supergravity description, a limitation to be extended also to non-perturbative contributions.  The idea of the present analysis is to use (exactly solvable) rational CFT models in string perturbation theory to explore special corners of the string theory "landscape" where the vacua exhibit a very small number of moduli.  The tools we use are non-geometric (chiral) twists and shifts on rational tori.  The hope, as discussed for instance in Ref.~\citen{triado}, is that a still unknown vacuum selection principle forces Nature to prefer these rather unconventional corners of the moduli space, where one can gauge and twist symmetries that are present only at special values of the parameters\cite{asym,special}. Asymmetric chiral twists freeze untwisted moduli, while freely-acting non-geometric shifts dispose of twisted moduli.  What is reported here is based on Ref.~\citen{noi}, where the interested reader can find more details and a more complete list of references.

\section{Four-dimensional Models with $\cN=1_L + 1_R$}

We scan two classes of models. The first one is a $\Z_{2L}\sigma_A \times
\Z_{2L}' \sigma_B \times \Z_{2R}\bar{\sigma}_C\times \Z_{2R}'\bar{\sigma}_D$ orbifolds
of Type IIB on the maximal $T^6$ torus
of $SO(12)$, where $\Z_{2}$'s are chiral inversions $I$ and $\sigma$'s are half-shifts in specific directions.  We use the description in terms of free fermions\cite{freef}.  Thus, each vacuum is specified by the boundary conditions assigned to the fermions and corresponds to a certain basis of fermionic sets containing the fermions odd under reflections. The starting torus corresponds to the sets
\be
F = \{ \psi^{1\ldots 8} \, y^{1\ldots 6} \,
w^{1\ldots 6}  | \, \tilde\psi^{1\ldots 8}\, \tilde{y}^{1\ldots 6}\,\tilde{w}^{1\ldots 6}   \} \ , \
S = \{\psi^{1\ldots 8}  \}\ , \ \tilde{S} = \{\tilde
\psi^{1\ldots 8}  \} \ , \ee
while the orbifolds are specified by four additional sets
\bea &&
b_1 =(b_{1L},b_{1R})= I_{3456}\, \sigma^{i_1 i_2 \ldots }\,\bar \sigma^{k_1 k_2 \ldots } = \{ (\psi \,y)^{3456}  \, (y\, w)^{i_1 i_2 \ldots }  | (\tilde y\, \tilde w)^{k_1 k_2 \ldots }  \} \ , \nn\\
&&b_2 =  (b_{2L},b_{2R})=I_{1256}\, \sigma^{j_1 j_2 \ldots }\,\bar \sigma^{l_1 l_2 \ldots } = \{(\psi\, y)^{1256}  \, (y\, w)^{j_1 j_2 \ldots }  | (\tilde y\, \tilde w)^{ l_1 l_2 \ldots }   \} \ ,
\eea
and the mirrors with left and right parts exchanged.  The scan is performed over the possible choices of the indices
$(i_1  i_2 \ldots)$, $(j_1 j_2 \ldots)$,   $(k_1  k_2 \ldots)$,  $(l_1 l_2 \ldots)$, compatibly with the conditions of modular invariance respecting spin-statistics
\be
   n(b_\alpha)=0~{\rm mod }~8 \ , \ n(b_\alpha \cap b_\beta)=0~{\rm mod }~4 \ , \
   n(b_\alpha \cap b_\beta\cap b_\gamma \cap b_\sigma)=0~{\rm mod }~2 \ ,   \label{consistency}
   \ee
where $n(b_\alpha)$ is the difference between the number of Left- and Right- fermions in the set $b_\alpha$.
The resulting three series of $\cN =1_L + 1_R$ susy models are reported in Table 1.  The spectra can be described in terms of ``effective'' Hodge numbers $h_{11}=n_h-1$ and $h_{21}=n_v$, in the spirit of geometrical compactifications. In Table 1 are also reported the breakings of the $SO(12)$ symmetry and the Euler characteristics $\chi$, always multiple of $12$.  It should be noted that the model with $(h_{11}=1,h_{21}=1)$ possesses the minimal massless content ever built in string perturbation theory\cite{altri}.

\section{Four-dimensional Models with $\cN=1_L$}

The second class of models is a genuinely asymmetric orbifold\cite{asym} of Type IIB to produce
$\cN =1_L$ spacetime supersymmetric vacua.  The GSO projection is performed only on the left modes and corresponds to the inclusion of the sets $F$, $S$, besides the scanning on the $b_1$ and $b_2$ (not the mirrors) on the lines of the previous Section.  The results are reported in Table 2, where the unprimed multiplets come fron the NS-NS sector while the primed ones from the R-R sector.  The tachyonic vacuum in the right sector makes the reduction of moduli less significant than in the left-right symmetric case.

\section{Four-dimensional Type I Models}

It would be interesting to analyze all the Type I descendants\cite{as} of the models in Section 2.  Unfortunately, the general case is very complicated due the large number of characters involved in the CFT description.  We limit ourselves to just an example: the unoriented projection of the $(1,1)$ model. It results in a vacuum without open strings with $\cN=1$ supergravity and only $2$ chiral massless multiplets.  Other examples with open strings can be found in Ref.~\citen{noi}.  Unfortunately, all the obtained models contain few moduli but are non-chiral, suggesting a possible tension between chirality and moduli stabilization\cite{ambt,tension}.

\section*{Acknowledgements}
It is a pleasure to thank the Organizers of the MG12 meeting for the beautiful conference and the very stimulating atmosphere, and in particular J.W. van Holten and P. Vanhove for having given to me the opportunity to present this
work.

\begin{center}
\begin{sidewaystable}
\tbl{Scan results for the $\cN =1_L + 1_R$ Type IIB Four-dimensional Models}{
\begin{tabular}{||c||c|c|c|c||}\hline
{\rm Models}&{$(h_{1,1}, h_{1,2})$}&{$b_1$}&{$b_2$}&{$SO(12)$}\\
\hline\hline
{$(n,n)$}&{$(1,1)$}&{$I_{3456} \ \sigma_1 \ \bar\sigma_5$}&{$I_{1256} \ \sigma_3 \ \bar\sigma_{12345}$} &{$SO(2)^4 \ \times \ O(1)^4$}\\ \cline{2-5}
{}&{$(2,2)$}&{$I_{3456} \ \sigma_1 \ \bar\sigma_2$}&{$I_{1256} \ \sigma_3 \ \bar\sigma_{12345}$} &{$SO(3) \ \times SO(2)^2 \ \times \ O(1)^5$}\\ \cline{2-5}
{}&{$(3,3)$}&{$I_{3456} \ \sigma_{12} \ \bar\sigma_{123456}$}&{$I_{1256} \ \sigma_{236} \ \bar\sigma_{1}$} &{$SO(3)^2 \ \times SO(2)^2 \ \times \ O(1)^2$}\\ \cline{2-5}
{$\chi=0$}&{$(4,4)$}&{$I_{3456} \ \sigma_{12} \ \bar\sigma_{5}$}&{$I_{1256} \ \sigma_3 \ \bar\sigma_{12456}$} &{$SO(3) \ \times SO(2)^2 \ \times \ O(1)^5$}\\ \cline{2-5}
{}&{$(5,5)$}&{$I_{3456} \ \sigma_{126} \ \bar\sigma_{12}$}&{$I_{1256} \ \sigma_{346} \ \bar\sigma_{35}$} &{$SO(2)^4 \ \times \ O(1)^4$}\\ \cline{2-5}
{}&{$(9,9)$}&{$I_{3456} \ \sigma_{12} \ \bar\sigma_{12}$}&{$I_{1256} \ \sigma_{34} \ \bar\sigma_{56}$} &{$SO(2)^6$}\\
\hline\hline
{$(2n,2n+6)$}&{$(0,6)$}&{$I_{3456} \ \sigma_{12} \ \bar\sigma_{15}$}&{$I_{1256} \ \sigma_{34} \ \bar\sigma_{36}$} &{$SO(2)^3 \ \times \ O(1)^6$}\\ \cline{2-5}
{}&{$(2,8)$}&{$I_{3456} \ \sigma_1 \ \bar\sigma_4$}&{$I_{1256} \ \sigma_{356} \ \bar\sigma_{2}$} &{$SO(3)^2 \ \times SO(2) \ \times \ O(1)^4$}\\ \cline{2-5}
{$\chi=-12$}&{$(4,10)$}&{$I_{3456} \ \sigma_1 \ \bar\sigma_5$}&{$I_{1256} \ \sigma_{346} \ \bar\sigma_{25}$} &{$SO(3)^2 \ \times SO(2) \ \times \ O(1)^4$}\\ \cline{2-5}
{}&{$(6,12)$}&{$I_{3456} \ \sigma_1 \ \bar\sigma_{12456}$}&{$I_{1256} \ \sigma_{356} \ \bar\sigma_{23456}$} &{$SO(3)^2 \ \times SO(2) \ \times \ O(1)^4$}\\ \cline{2-5}
\hline
{$(2n+6,2n)$}&{$(6,0)$}&{$I_{3456} \ \sigma_{12} \ \bar\sigma_{13}$}&{$I_{1256} \ \sigma_{34} \ \bar\sigma_{25}$} &{$SO(2)^3 \ \times \ O(1)^6$}\\ \cline{2-5}
{}&{$(8,2)$}&{$I_{3456} \ \sigma_1 \ \bar\sigma_2$}&{$I_{1256} \ \sigma_{356} \ \bar\sigma_{4}$} &{$SO(3)^2 \ \times SO(2) \ \times \ O(1)^4$}\\ \cline{2-5}
{$\chi=12$}&{$(10,4)$}&{$I_{3456} \ \sigma_{12} \ \bar\sigma_{45}$}&{$I_{1256} \ \sigma_{36} \ \bar\sigma_{5}$} &{$SO(3)^2 \ \times SO(2) \ \times \ O(1)$}\\ \cline{2-5}
{}&{$(12,6)$}&{$I_{3456} \ \sigma_1 \ \bar\sigma_{23456}$}&{$I_{1256} \ \sigma_{356} \ \bar\sigma_{12456}$} &{$SO(3)^2 \ \times SO(2) \ \times \ O(1)^4$}\\ \cline{2-5}
\hline\hline
{$(2n+3,2n+15)$}&{$(3,15)$}&{$I_{3456} \  \bar\sigma_{3456}$}&{$I_{1256} \  \bar\sigma_{1256}$} &{$SO(2)^6$}\\ \cline{2-5}
{$\chi=-24$}&{$(5,17)$}&{$I_{3456} \ \sigma_{12} \ \bar\sigma_{34}$}&{$I_{1256} \ \sigma_{34} \ \bar\sigma_{123456}$} &{$SO(4) \ \times SO(2)^4$}\\ \cline{2-5}
\hline
{$(2n+15,2n+3)$}&{$(15,3)$}&{$I_{3456} \ \sigma_{12} \ \bar\sigma_{12}$}&{$I_{1256} \ \sigma_{34} \ \bar\sigma_{34}$} &{$SO(2)^6$}\\ \cline{2-5}
{$\chi=24$}&{$(17,5)$}&{$I_{3456} \ \sigma_{126} \ \bar\sigma_{123456}$}&{$I_{1256} \ \sigma_5 \ \bar\sigma_{3456}$} &{$SO(4) \ \times SO(2)^4$}\\ \cline{2-5}
\hline
\end{tabular}}
\vskip 36pt
\tbl{Scan results for the $\cN =1_L$ Type IIB Four-dimensional Models}{
\begin{tabular}{||c|c|c|c||}\hline
{$(n_v,n_{v\prime};n_c,n_{c\prime})$}&{$b_1$}&{$b_2$}&{$SO(12)_L \times SO(20)_R$}\\
\hline\hline
{$(14,0;5,0)$}&{$I_{3456} \ \sigma_{12} \ \bar\sigma_{45}$}&{$I_{1256} \ \sigma_{36} \ \bar\sigma_{5}$} &{$[ SO(4)^2 \ \times \ SO(2)^2]_L \times [ SO(16) \times SO(2)^2]_R$}\\ \hline
{$(10,0;25,0)$}&{$I_{3456} \ \sigma_{126} \ \bar\sigma_{12}$}&{$I_{1256} \ \sigma_{346} \ \bar\sigma_{35}$} &{$[ SO(6) \ \times \ SO(2)^3]_L \times [ SO(4)^2 \times SO(12)]_R$}\\ \hline
{$(8,0;27,0)$}&{$I_{3456} \ \sigma_{1} \ \bar\sigma_{5}$}&{$I_{1256} \ \sigma_{3} \ \bar\sigma_{12345}$} &{$[ SO(4)^2 \ \times \ SO(2)^2]_L \times [ SO(8) \times SO(2) \times SO(10)]_R$}\\ \hline
{$(6,8;13,8)$}&{$I_{3456} \ \sigma_{12} \ \bar\sigma_{45}$}&{$I_{1256} \ \sigma_{36} \ \bar\sigma_{5}$} &{$[ SO(4)^2 \ \times \ SO(2)^2]_L \times [ SO(16) \times SO(2)^2]_R$}\\ \hline
{$(6,8,29,8)$}&{$I_{3456} \ \sigma_{12} \ \bar\sigma_{45}$}&{$I_{1256} \ \sigma_{36} \ \bar\sigma_{5}$} &{$[ SO(4)^2 \ \times \ SO(2)^2]_L \times [ SO(16) \times SO(2)^2]_R$}\\ \hline
\end{tabular}}
\label{tab2}
\end{sidewaystable}
\end{center}

\end{document}